\documentstyle[]{mn}

\newif\ifAMStwofonts
\ifoldfss
  \ifCUPmtlplainloaded \else
    \NewTextAlphabet{textbfit} {cmbxti10} {}
    \NewTextAlphabet{textbfss} {cmssbx10} {}
    \NewMathAlphabet{mathbfit} {cmbxti10} {} % for math mode
    \NewMathAlphabet{mathbfss} {cmssbx10} {} %  "   "    "
  \fi
  \ifAMStwofonts
    \ifCUPmtlplainloaded \else
      \NewSymbolFont{upmath} {eurm10}
      \NewSymbolFont{AMSa} {msam10}
      \NewMathSymbol{\upi}     {0}{upmath}{19}
      \NewMathSymbol{\umu}     {0}{upmath}{16}
      \NewMathSymbol{\upartial}{0}{upmath}{40}
      \NewMathSymbol{\leqslant}{3}{AMSa}{36}
      \NewMathSymbol{\geqslant}{3}{AMSa}{3E}

      \let\leq=\leqslant 
       
    \fi
  \fi
\fi % End of OFSS

\ifnfssone
  \newmathalphabet{\mathit}
  \addtoversion{normal}{\mathit}{cmr}{m}{it}
  \addtoversion{bold}{\mathit}{cmr}{bx}{it}
  \newmathalphabet{\mathbfit} % math mode version of \textbfit{..}
  \addtoversion{normal}{\mathbfit}{cmr}{bx}{it}
  \addtoversion{bold}{\mathbfit}{cmr}{bx}{it}
  \newmathalphabet{\mathbfss} % math mode version of \textbfss{..}
  \addtoversion{normal}{\mathbfss}{cmss}{bx}{n}
  \addtoversion{bold}{\mathbfss}{cmss}{bx}{n}
  \ifAMStwofonts
    \ifCUPmtlplainloaded \else
      \UseAMStwoboldmath
      \makeatletter
      \new@mathgroup\upmath@group
      \define@mathgroup\mv@normal\upmath@group{eur}{m}{n}
      \define@mathgroup\mv@bold\upmath@group{eur}{b}{n}
      \edef\UPM{\hexnumber\upmath@group}
      \new@mathgroup\amsa@group
      \define@mathgroup\mv@normal\amsa@group{msa}{m}{n}
      \define@mathgroup\mv@bold\amsa@group{msa}{m}{n}
      \edef\AMSa{\hexnumber\amsa@group}
      \makeatother
      \mathchardef\upi="0\UPM19
      \mathchardef\umu="0\UPM16
      \mathchardef\upartial="0\UPM40
      \mathchardef\leqslant="3\AMSa36
      \mathchardef\geqslant="3\AMSa3E

      \let\leq=\leqslant 

    \fi
  \fi
\fi % End of NFSS release 1

\ifnfsstwo
  \DeclareMathAlphabet{\mathbfit}{OT1}{cmr}{bx}{it}
  \SetMathAlphabet\mathbfit{bold}{OT1}{cmr}{bx}{it}
  \DeclareMathAlphabet{\mathbfss}{OT1}{cmss}{bx}{n}
  \SetMathAlphabet\mathbfss{bold}{OT1}{cmss}{bx}{n}
  \ifAMStwofonts
    \ifCUPmtlplainloaded \else
      \DeclareSymbolFont{UPM}{U}{eur}{m}{n}
      \SetSymbolFont{UPM}{bold}{U}{eur}{b}{n}
      \DeclareSymbolFont{AMSa}{U}{msa}{m}{n}
      \DeclareMathSymbol{\upi}{0}{UPM}{"19}
      \DeclareMathSymbol{\umu}{0}{UPM}{"16}
      \DeclareMathSymbol{\upartial}{0}{UPM}{"40}
      \DeclareMathSymbol{\leqslant}{3}{AMSa}{"36}
      \DeclareMathSymbol{\geqslant}{3}{AMSa}{"3E}

      \let\leq=\leqslant 

    \fi
  \fi
\fi % End of NFSS release 2

\ifCUPmtlplainloaded \else
  \ifAMStwofonts \else % If no AMS fonts
    \def\upi{\pi}
    \def\umu{\mu}
    \def\upartial{\partial}
  \fi
\fi

\title{Searching for Stars in Compact High-Velocity Clouds. I First Results from VLT and 2MASS}
\author[Hopp et al.]
       {U. Hopp$^1$,\thanks{\sf Based on observations collected 
at the European Southern Observatory, Chile, during run 67.B-0060(A).}
       R.E. Schulte-Ladbeck$^2$,
       J. Kerp$^3$ \\
       $^1$Universit\"ats-Sternwarte M\"unchen, Scheinerstr. 1, D-81679 M\"unchen, Germany\\
       $^2$University of Pittsburgh, Department of Physics \& Astronomy, Pittsburgh, PA 15260, USA\\
       $^3$Radioastronomisches Institut der Universit\"at Bonn, Auf dem H\"ugel 71, D-53121 Bonn, Germany}
\date{Accepted .
      Received ;
      in original form }

\pagerange{\pageref{firstpage}--\pageref{lastpage}}
\pubyear{2002}

\begin{document}

\maketitle

\label{firstpage}

\begin{abstract}

We investigate the hypothesis that compact high-velocity clouds (CHVC)
are the ``missing" dwarf galaxies of the Local Group, by searching them
for populations of resolved stars. To this end we conducted two distinct
tests based on optical and near-infrared single-star photometry.
The optical and the near-infrared experiments complement one another; the
optical data help us to rule out distant populations but they are restricted to
the central regions of the gas distributions, whereas the near-infrared 
photometry allows us to set limits on nearby populations spread over the typical
cloud size. First, we discuss deep optical single-star
photometry of five CHVCs in the V and I filters, obtained with the FORS instrument
at the Very Large Telecope (VLT). We find that their optical colour-magnitude diagrams
are indistinguishable from that of a population of Galactic stars, and attribute
all of the resolved stars to Galactic foreground. We present simulations
which address the question of how much of a ``normal" dwarf-galaxy type population
we might hide in the data. A Kolmogorov-Smirnov test allows us to set very stringent
limits on the absence of a resolved stellar population in CHVCs. Second, we also
culled near infrared single-star photometry in the J, H, and K$_S$ bands for four of the CHVCs from
the Two Micron All Sky Survey (2MASS). The infrared
data do not reveal any stellar contents in the CHVCs which resembles that of 
nearby dwarf galaxies either, and are explained with Galactic foreground 
as well. We interpret our null detections to indicate that the five CHVCs investigated 
by us do not host an associated stellar content which is similar to that
of the known dwarf galaxies of the Local Group. These CHVCs are very likely
pure hydrogen clouds in which no star formation has taken place over cosmic time.

\end{abstract}


\section{Introduction}

Cold Dark Matter (CDM) cosmogonies foretell the existence of a significant 
amount of substructure in galaxy halos (Klypin et al. 1999, Moore et al. 1999). 
However, the frequency of low-mass dark-matter halos around the Milky Way predicted in
numerical simulations of structure formation, is an order of magnitude higher than 
the known number of dwarf galaxies in the Local Group (Mateo 1998). 
Theoretical solutions of this CDM ``crisis" include propositions to change the properties
of the dark matter particle (e.g., Spergel \& Steinhardt 2000; Col\'in, Avila-Reese \&
Valenzuela 2000), as well as proposals which seek to change with cosmic time the astrophysical conditions 
that make gas conducive to star formation in small halos (e.g. Bullock, Kravstov \& Weinberg 2001).
An alternative solution is the hypothesis that the subhalos predicted by CDM simulations
have been overlooked so far observationally. A potential source for these ``missing" dwarf 
systems is the large population of Compact High-Velocity Clouds (CHVC). 

\begin{table*}
 \centering
 \begin{minipage}{140mm}
  \caption{Compact High Velocity Clouds Observed With VLT/FORS.}
  \begin{tabular}{lrrrrr}
Name &  RA [h:m:s]  & Dec [$^o$:':"] & A$_V$ [m] & A$_K$ [m] & N stars\\
HIPASS~J1712-64 & 17:12:35 & -64:38:00 & 0.352 & 0.039 & 3931\\
HVC032-31-299   & 20:42:46 & -14:15:10 & 0.142 & 0.016 & 865\\
HVC039-31-265   & 20:53:13 & -08:46:56 & 0.240 & 0.027 & 999\\
HVC039-33-260   & 21:01:05 & -10:20:10 & 0.393 & 0.044 & 732 \\
HVC039-37-231   & 21:16:25 & -11:43:44 & 0.140 & 0.016 & 1104\\ 

\end{tabular}
\end{minipage}
\end{table*}

\begin{figure*} 
%\centerline{\psfig{figure=hvc_fig1.ps,width=15.0cm,angle=0}} 
\caption{CMDs for four CHVCs with similar Galactic latitudes. 
Only the photometry of the sharpest images is plotted (fewer points than N~stars in Table~1).
The morphology of
the CMDs is indistinguishable from that of a population of Galactic stars. 
Notice the sharp edge at V-I$\approx$1, blueward of which we do not observe any stars.
There is no ``blue plume" of supergiants and main-sequence stars which is characteristic of
the CMDs of star-forming galaxies of the Local Group. There is also no distinctive
``red plume" of red supergiants, asymptotic giants, and red giant stars. The RGB in 
particular is a prominent feature on the CMDs of all dwarf galaxies of the Local Group.
We also overplot isochrones from Girardi et al. (2000) for a distance of 1~Mpc,
Z=0.004 (1/5 of Solar) and ages of 10~Myr, 100~Myr, 1~Gyr, and 10~Gyr. This outlines
the regions where the [V-I, I] CMD of an associated galaxy is expected to 
be populated with stars.
 \label{figlabel}}
\end{figure*}

High-Velocity Clouds are concentrations of neutral hydrogen with extremely high radial velocities that 
are inconsistent with Galactic rotation models. In spite of decades of intense investigation, the nature
of HVCs remains unknown. In their seminal review, Wakker \& van Woerden (1997) distinguished three 
different origins of HVCs: Galactic fountain, particularly for those clouds seen in absorption against 
halo stars within the Milky Way, the tidal debris of the Magellanic Stream, and ``others". 

Blitz et al. (1999) proposed that some of the ``other" HVCs could be distant, extragalactic HI clouds 
distributed throughout the Local Group. They suppose such HVCs could in fact represent primordial
gas clouds or protogalaxies, which could be interpreted as the leftover building blocks from 
which Local Group galaxies have formed. These HVCs are compact, and isolated from the large 
HVC complexes. Blitz et al. suggest that these compact and isolated HVCs, have a characteristic 
mean distance of about 1~Mpc. Do compact, isolated HVCs trace the
substructures predicted by CDM 
simulations?

Braun \& Burton (1999, 2000) identified an intial catalog of 65 Compact HVCs, and conducted detailed 
follow-up observations. CHVCs have a mean infall velocity of 100~km~s$^{-1}$ in the Local Group 
reference frame, have angular sizes less than 2$^o$~FWHM, are isolated, and, at an {\it assumed} 
distance of 1~Mpc, have HI masses of a few times 10$^7$~M$_{\odot}$. Their rotation curves 
imply a high dark-to-visible mass ratio of 10 to 50. Deep integrated HI maps reveal a 
core-halo structure (Br\"uns, Kerp \& Pagels 2001). 
The highest column density and coldest HI gas is situated in a compact region no more than several 
arcmin across, whereas the large-scale gas distribution consists of warm, lower column density gas. 
The cold, dense regions, would be the regions which are most conducive to star formation. 
The HI column densities of these regions are in excess of N$_{HI}$$>$3$\times$10$^{20}$~cm$^{-2}$
(Herbstmeier, Heithausen, and Mebold 1993), 
sufficiently high to form molecules and star-forming cores. Putman et al. (2002) recently identified
additional CHVCs in the HI Parkes All-Sky Survey (HIPASS). There are 179 objects on the southern
hemipshere which share many of the properties of the original CHVCs identified by Braun \& Burton. 
Additionally, de Heij, Braun \& Burton (2002) re-analyzed the Leiden/Dwingeloo Survey data 
from which the original Braun \& Burton list was drawn, and recognise 67
CHVCs. The current source count of CHVCs thus stands at 246.

A major hurdle in understanding the nature of HVCs in general, and that of CHVCs in
particular, has been the lack of distance information. Braun (2001) recently reviewed existing, 
and hitherto unsuccessful, attempts to derive distances. The goal of identifying
stars associated with CHVCs is that stars would supply distance indicators. Yet, CHVCs have no optical 
counterparts on sky survey plates; this rules out resolved stars down to 
V$\approx$21. An associated content of bright young supergiants seems highly unlikely.
Simon \& Blitz (2002) recently inspected digitized Palomar Sky Survey (POSS) plates
with a spatial filtering method, for 264 northern HVCs. The surface brightness limits reached
by their analysis (26 magnitudes~arcsec$^{-2}$ in V) would have recovered all known Local Group 
galaxies except for four of the very diffuse, extended dwarf Spheroidal (dSph) galaxies within 
100~kpc of the Milky Way. Follow-up imaging of several ``suspicious" CHVCs at 1-m and 3-m telescopes 
did not reveal a stellar content to a typical limiting stellar magnitude of R$_S$ = 22.2.
Simon \& Blitz rule out the possibility that HVCs are associated with a sub-population of dwarf galaxies
within the range of known dwarf galaxy properties.
Other attempts to resolve a faint stellar content in CHVCs with 4-m class telescopes have failed due to 
crowding of potential stars in CHVCs with foreground stars and background galaxies at the arcsec level,
although Braun (2001) reports there is tantalizing evidence of the possible detection of 
tip-of-the-red-giant-branch (TRGB) stars in a few, which awaits confirmation. 

We here present two new and complementary searches for the elusive stellar content of CHVCs. The
first approach uses deep imaging in the optical with the 8.2-m Very Large Telescopes
(VLT). The data are presented and discussed in section~2. The VLT data consist of optical imaging with a
reasonably sized field of view, 6\farcm8 x 6\farcm8, centered on the highest column density
regions of five CHVCs. We show that these observations have sufficiently high angular resolution
and limiting magnitude to be sensitive to red giant branch stars (the brightest 
phase of a 1--12~Gyr stellar population), throughout the Local Group. 
The second is near-infrared single-star photometry using 2MASS data. We selected all stars
within a radius of 1$^o$ of the CHVC positions for analysis. The data are presented
and discussed in section~3. We show that the 2MASS data are sensitive to intermediate and
old stellar populations in the dwarf Spheroidal companions of the Milky Way.

Since our observations do not reveal a stellar content in five HVCs, we conclude in section~4
that the five objects studied by us did not experience any star formation over cosmic time.
Even with such a small sample, we can already place some
interesting constraints on the putative stellar content of HVCs.

\section{VLT Data}

We submitted a target list of southern CHVCs for VLT service observing to the good, 
$<$0\farcs6, seeing queue. Here, we report on the results of our first semester of observations, with
a fall right ascension coverage. The results of VLT service mode observations to be 
conducted in the spring of 2003 will be the subject of our paper~II.

In order to minimise contamination of the images by foreground stars in the Galaxy, we chose 
targets at Galactic latitudes above $|$30$^o$$|$. The target list was drawn from Braun \& Burton (1999) 
and emphasises CHVCs observed with radio interferometers from Braun \& Burton (2000), 
Br\"uns et al. (2000), and from unpublished HI observations by one of us (Kerp). We added to our
target list, HIPASS~J1712-64. Kilborn et al. (2000) drew attention to this isolated, star-free cloud of
neutral hydrogen, which they report could be at a distance of 3.2~Mpc (based on its HI velocity and
Hubble's law).

\begin{figure} 
%\centerline{\psfig{figure=hvc_fig2.ps,width=8.0cm,angle=0}} 
\caption{CMD for HIPASS~J1712-64. Notice the sharp edge at V-I$\approx$1, blueward
of which we do not observe ``any" stars. There is no young population
in HIPASS~J1712-64 (compare with the main-sequence of Phoenix, a distinct
feature in Fig.~4 at a V-I colour of about 0). 
We also overplotted Globular Cluster ridgelines from Da Costa
\& Armandroff (1990) for a distance of 3.2~Mpc. This indicates that such a hypothetical RGB population 
is near the detection limit of our data. 
 \label{figlabel}}
\end{figure}

\begin{figure} 
%\centerline{\psfig{figure=hvc_fig3.ps,width=7.0cm,angle=0}} 
\caption{The DAOPHOT errors for all five CHVCs. The data quality is very good,
with errors smaller than 0.1 to an I magnitude of about 24 and a V magnitude of
about 25.5. (The few outliers at the bright end arise from  
bright stars near the saturation limit of the CCD.)
 \label{figlabel}}
\end{figure}

\begin{figure} 
%\centerline{\psfig{figure=hvc_fig4.ps,width=8.0cm,angle=0}} 
\caption{The CMD of the ``transition" type dwarf Irregular/Spheroidal galaxy Phoenix,
which has a distance of about 0.45~Mpc.
 \label{figlabel}}
\end{figure}

\begin{figure} 
%\centerline{\psfig{figure=hvc_fig5.ps,width=7.0cm,angle=0}} 
\caption{The DAOPHOT errors for Phoenix. The data quality is similar to that of the
five CHVCs (cf. Fig.~3), in the sense that errors remain smaller than 0.1 to an I magnitude 
of about 24 and a V magnitude of about 25.5.
 \label{figlabel}}
\end{figure}

We obtained observations of five HVCs (including HIPASS~J1712-64). Details of the targets
are presented in Table~1. The coordinates reflect the J2000 centres of our VLT poitings. 
According to the naming convention of Braun \& Burton (1999), the name consists of
a three-digit Galactic longitude, followed by a two-digit Galactic latitude, and a three-digit
local-standard-of-rest velocity. The Galactic coordinates for HIPASS~J1712-64 are l$\approx$326\fdg5,
b$\approx$-14\fdg5. Notice that this cloud is at a lower Galactic latitude than the
remainder of the sample.
The Galactic absorptions were obtained using NED\footnote{The NASA/IPAC Extragalactic Database 
(NED) is operated by the Jet Propulsion Laboratory, California Institute of Technology,
under contract with the National Aeronautics and Space Administration.}. NED provides Galactic
extinctions from Schlegel, Finkenbeiner \& Davis (1998), and assumes a Galactic extinction law 
with R$_V$=3.1 (Cardelli, Clayton \& Mathis 1989).

The observations were carried out at the VLT UT1 ``ANTU" with the FORS1 instrument. They
consisted of dithered exposures in I and V, for each of the
CHVCs. The intregration times measured 5~$\times$~300s in I and 3~$\times$~300s in V, except for
HVC039-33-260, for which they were 5~$\times$~200s in I and 3~$\times$~200s in V.
Each observation was accompanied by a quality file. We checked that the conditions during
the observations were photometric, and that the seeing was no worse than 0\farcs6. We used the 
pipeline-reduced data in our analysis. The individual exposures were shifted and added. 
The I-band data needed additional flat-field correction beyond that already applied in the
pipeline. Single-star photometry was carried out with
DAOPHOT~II (Stetson 1992). We used isolated stars in each field in order to determine 
the point-spread function (PSF).
We find that the FWHM was between 0\farcs52 and 0\farcs8, with a mean of 0\farcs65.
Photometric zero points for each night and in each filter were obtained from the European Southern
Observatory's (ESO) homepage. Extinction terms in V and I, and V-I colour terms needed
to transform to the Johnson-Cousins photometric system, were
gleaned from the ESO homepage as well. We used the PSF stars to find the zero points, determine
aperture corrections, and define the transformation of the data into the Johnson-Cousins V, I system.
The DAOPHOT parameters sharpness and roundness were used to clip extended sources, i.e.,
mainly background galaxies, with sizes larger than the PSF from the data tables. The last column
in Table~1 (N stars) lists for each CHVC, the number of objects classed to be stars and detected in 
V and I.

Our data achieve limiting magnitudes in V$\approx$25--26, and in I$\approx$23--24. In Figure~1,
we display the [V-I,~I] CMDs of the four CHVCs located at similar Galactic latitudes (see Table~1).
In Figure~2, we show the [V-I,~I] CMD of HIPASS~J1712-64. As an assessment of the data quality we
reproduce in Figure~3 the DAOPHOT photometric errors in the V \& I bands. We overplotted the errors for 
all five fields, since the data were obtained with the same (or similar) exposure times and have 
similar properties.

\subsection{CHVCs versus Local Group Dwarf Galaxies}

We have no a priori knowledge of the star-formation history (SFH) of CHVCs. 
CHVCs could have just started to form stars,
and then their CMDs would exhibit nothing but the ``blue plume". This seems somewhat unlikely given the non-detection
of stellar content on survey plates. If CHVCs started to form stars
more than a Gyr ago, their CMDs should exhibit the ``red plume".  Indeed, the dwarf galaxies of
the Local Group, which experienced a variety of SFHs, always show stars on the red plume. 
The CMDs of Local Group dwarfs exhibit either the red plume only (dSph), 
a weak blue plume as well as a red plume (dIrr/dSph), or
a strong blue plume combined with a red plume (dIrr). These morphologies serve as a 
guide to our exploration of CHVC CMDs. 

We display in Figure~4 the [V-I,~I] CMD of the transition dIrr/dSph galaxy
Phoenix. The observations were obtained with the VLT UT2 ``KOYEN" and FORS2 under good 
seeing conditions. 
The exposure times were 3~$\times$~300s each, in V and I. The data
were reduced, and single-star photometry applied in the same way, as described above
for the CHVCs. We measure a FWHM in V of 0\farcs54, and in I of 0\farcs7.
The total number of stars detected in V and I is 8780, much higher than
in any of the CHVC fields. The DAOPHOT photometric errors are displayed in Figure~5, in order
to emphasise that the quality of these data is similar to that of the CHVC data sets.
Phoenix is located at high Galactic latitude, near --69$^o$. The foreground contamination
and extinction are small. It has a 
distance modulus of about 23.25 (Mateo 1998) or a distance of about 450~Kpc. The CMD of
Phoenix shows why it is classed as having a mixed morphology: it exhibits both a blue plume
of relatively young stars, and a red plume which includes the RGB. 

The stellar distributions on the CMDs of the CHVCs are quite different from that of known
Local Group dwarfs. The morphology of the CHVC CMDs does not reveal either of the two plumes. 
Instead, the CMDs of the CHVCs are reminiscent of other, deep HST pointings
into the disk and halo of our Galaxy (e.g., Richer et al. 2002).

We overplotted on Fig.~1 theoretical isochrones from the database of
Girardi et al. (2000) for ages of 10 and 100~Myr, and for 1 and 10~Gyr. We
adopted  
a metallicity of 1/5 of Solar, as dwarf galaxies tend to have sub-Solar
metallicities. In general, the effect of varying metallicity at a given age is
that more metal-poor isochrones shift to bluer V-I colors, 
while more metal-rich ones show redder colors. This shift
in minimal for the ``young" isochrones (e.g., 10 \& 100~Myr), but noticable for
the ``old" isochrones (1 \& 10~Gyr). We show the set with a metallicity of 1/5 of
Solar because the 1 \& 10~Gyr isochrones have sufficiently red colors to 
encompass the data distribution.

When comparing the data with the ``young" isochrones, we notice there are
virtually no stars observed with V-I$<$1, where we expect to find a young population. 
This emphasises the absence of young MS and blue supergiant stars in the CHVCs. 

The brightest evolutionary phase of stars with ages larger than
about one Gyr occurs on the red giant branch. Furthermore, the TRGB, at M$_I$ of 
--4, is a well calibrated distance indicator (Lee, Freedman \& Madore 1983), and at a distance of 1~Mpc,
it would appear at I$\approx$21. We overplotted on Fig.~1 theoretical isochrones with
ages of 1 and 10~Gyr. Note that the Girardi et al. isochrones extend above
the RGB into the thermally pulsing AGB phase. A TRGB would have been detected for the entire 
metallicity range appropriate for dwarf galaxies, 
to I magnitudes of at least 22.5, or to a distance of up to 2~Mpc, 
in the four CHVCs whose CMDs are shown in Fig.~1. 
Notice that the absence of stars with V-I$<$1 rules out extremely
metal-poor RGBs, which can be observed in dwarf Spheroidal galaxies. 
Since no RGB is obvious, a straightforward
conclusion is that there is none, and that these objects are not faint galaxies within the
Local Group.  

\subsection{HIPASS~J1712-64}

The absence of a TRGB with I$\leq$21 suggests the hypothesis that HIPASS~J1712-64 is a faint dwarf galaxy
of the Local Group is similarly unlikley. To illustrate the expected locations of RGBs in our data 
if HIPASS~J1712-64 were at a distance of 3.2~Mpc as proposed by Kilborn et al. (2000), 
we overplotted in Fig.~2 the shifted Globular Cluster
ridgelines from da Costa \& Armandroff (1990) of M~15, [Fe/H]=-2.17, NGC~6397, -1.91, M~2, 
-1.58, NGC 6752, -1.54, NGC~1851, -1.29, and 47~Tuc, -0.71. Even if HIPASS~J1712-64 were as far as 3.2~Mpc, 
with a corresponding TRGB I-magnitude of 23.5, we could have detected it if its metallicity were low enough,
although this is clearly at the very limit of our data. Lewis et al. (2002) recently 
presented single-star photometry obtained under sub-arcsec seeing conditions 
in a 0\degr5$\times$0\degr5 field of view centered at 17~12~35.7, --64~38~12 (J2000) and 
reaching magnitudes fainter than r=24 and g=25. They, too,
failed to detect an obvious stellar content associated with HIPASS~J1712-64. Furthermore, they 
rule out any underlying unresolved diffuse components within their large field of view to deep
surface-brightness limits.

\begin{figure*} 
%\centerline{\psfig{figure=hvc_fig6_new.ps,width=15.0cm,angle=0}} 
\caption{The distribution of sources within a 2$^o$ field centered on the four CHVCs
for which 2MASS data are available. There
are no enhancements in stellar density toward the centre which would suggest the presence of a galaxy
associated with one of the clouds. (Only every other
data point is plotted for clarity.)
 \label{figlabel}}
\end{figure*}

\begin{figure*} 
%\centerline{\psfig{figure=hvc_fig7_new.ps,width=15.0cm,angle=0}} 
\caption{The distribution of sources within a 2$^o$ field in the comparison
fields at the same Galactic latitude as the respective CHVC but with longitudes centered
on l+3$^o$. The number of Galactic 
foreground sources is high in the near-infrared, even at Galactic latitudes above 30$^o$.
(Only every other data point is plotted for clarity.) 
\label{figlabel}}
\end{figure*}

\begin{figure*} 
%\centerline{\psfig{figure=hvc_fig8.ps,width=15.0cm,angle=0}} 
\caption{The [J-K$_S$,~K$_S$] CMDs of four CHVCs. The features seen in these
CMDs are as same as those seen in the corresponding OFF CMDs. We overplotted Girardi et al. (2000) isochrones
for a distance of 150~Kpc. As in Fig.~1, Z=0.004 (1/5 of Solar); and ages of 10~Myr, 100~Myr, 1~Gyr, and 10~Gyr
were selected. They indicate AGB star with colours $>$1.0 should appear in the data; none are seen. 
 \label{figlabel}}
\end{figure*}

\begin{figure*} 
%\centerline{\psfig{figure=hvc_fig9.ps,width=15.0cm,angle=0}} 
\caption{The [J-K$_S$,~K$_S$] CMDs of four satellite galaxies of the Milky Way. 
The LMC and the SMC are larger than the 2$^o$ field for which we
extracted data. In the case of Fornax and Sculptor, the galaxies are smaller than the 2$^o$ field
(see Fig.~10).
The black symbols plot stars within a 0.3$^o$ radius of the center of each galaxy,
while the bright dots correspond to the stars in the remainder of the
field. This allows us to crudely separate stars in the foreground from those
in Fornax and Sculptor. We overplotted Girardi et al. (2000) isochrones
for the respective distance of each galaxy. As before, we used a metallicity of Z=0.004 (1/5 of Solar), and ages of 
10~Myr, 100~Myr, 1~Gyr, and 10~Gyr. 
 \label{figlabel}}
\end{figure*}

\begin{figure*} 
%\centerline{\psfig{figure=hvc_fig10_new.ps,width=15.0cm,angle=0}} 
\caption{The distribution of sources within a 2$^o$ field centered on four dSph companions
of the Milky Way for which 2MASS data are available. The higher-surface-brightness galaxies
Fornax and Sculptor are easily distinguished from Galactic foreground stars, while the
lower-surface-brightness Ursa Minor and Sextans are not. (Only every other
data point is plotted for clarity.)
 \label{figlabel}}
\end{figure*}

\subsection{Detection Limits}

We can set upper limits on the non-detection of a possible CHVC stellar population in the 
VLT/FORS data, by constructing CMDs and luminosity functions (LF) with added sub-populations. 
This is accomplished by using the VLT FORS2 observation of Phoenix obtained under very
similar conditions and with similar photometric limits, as a template stellar
population with which we ``contaminate" the observed CHVC CMDs.

We added Phoenix stars to the HVC CMDs at their location in the Phoenix CMD. In other words, we
added the stars at the Phoenix distance. We did not vary the distance, since shifting the
data would require us to make a model for how to vary the errors and completeness
fractions accordingly. If we consider about 1~Mpc the radius of the Local
Group, than the Phoenix population, situated at a distance of 0.45~Mpc, 
yields an appropriate template at a typical, mean distance under the
hypothesis that CHVCs are dwarf galaxies scattered throughout the Local Group.

We reduced the number of stars in the Phoenix CMD before adding them
to the HVC CMDs. The number of stars was reduced by up to a factor of 200. Any young MS
population is easily detected (even with $\sim$20 stars only). This is because the
HVCs do not show any stars (or very few stars/noise spikes) with colours of the 
Phoenix MS. Because there are simply no stars at magnitudes where we would
expect to see a main-sequence, or a luminous blue supergiant stellar content in a [V-I, I]
colour magnitude diagram, it is even safe to say that if there had been a young stellar
population in the high column density regions of the HVC centres, then we would
have detected it out to the very limits of the Local Group ($\approx$~1~Mpc). 

A scarce RGB population is more difficult to detect. We therefore performed a one-dimensional 
KS test to compare the original and simulated RGB populations.
We constrained the test range to 18.5$<$I$<$22.25 and 0.5$<$V-I$<$1.5, as
expected for RGB stars. The HVCs typically show 100 to 150
stars in this colour and magnitude range, with the exception of HIPASS~J1712-64, which exhibits
477 stars. We obtained I-band LFs of the HVCs, and of the HVCs plus fractional
Phoenix RGB populations. We used the D-statistic to decide whether or not we detected the
population  of red giants in the simulated (original HVC plus fractional Phoenix) data as 
compared with the original HVC data. 

When the Phoenix RGB population is reduced to one fiftieth or less of its original strength 
(i.e., we are adding 26 or fewer stars from the Phoenix RGB), we cannot recover it. 
The probability that the original and the simulated populations are identical is equal 
to or above 95\% in all cases. We increased the strength of the Phoenix RGB in steps. 
Inspecting the simulated CMDs by eye, results in the recognition of dilute RGBs when
the strengths are between about 1/40 (30 stars) to 1/30 (56 stars) of the Phoenix RGB. 
At this level of ``contamination", the probability of the LFs being identical is still high, 
around 50\%. At 1/20 of its strength, we are adding 73 Phoenix RGB stars to the HVCs. The 
probability for the simulated and the original RGB populations to be identical is now small, 
less than or about 10\%. The added population is so obvious in the CMDs by eye, that it is likely that
we would indeed conclude to the presence of an RGB population.

These tests suggest up to about 70 RGB stars could comprise a hidden, undetected 
dwarf galaxy population among the foreground contamination within our field of view. 
In other words, we believe that we were sensitive to
detecting fewer than about 1.5 Phoenix-like RGB stars per arcmin$^2$, for
a distance of 0.5~Mpc. None are seen. We calculate that this corresponds to
a V-band surface brightness of 29 magnitudes arcsec$^{-2}$, or an M$_V$ of about
-5.8 (compare Simon \& Blitz 2002, Fig.~6).

So far, we analyzed data for five HVCs. We interpret our result to indicate
that we detect zero stellar content in five cases observed. We use the
binomial statistic to ask what is the probability of getting
zero successes in five trials. The hypothesis that all five HVCs contain stars which escaped us by chance,
can be ruled out with great confidence (the probability is smaller than 10$^{-8}$). The probability 
that half of the HVCs contains stars is also quite unlikely given our results, 3\%. 
There is, however, a 50\% probability of obtaining our result if the fraction of HVCs which 
contains stars is less than 13\%. As the sample of CHVCs searched for stellar content grows, these fractions
may change, and will become better defined.

The result of our VLT observations allow us to set rather strong limits 
on the absence of a dwarf-galaxy like stellar population for five CHVCs. 

\begin{enumerate}

\item Any young stellar 
population contributes less than a few percent (0.5 to 3\%) of the stars seen in the FORS 
field of view centered on HI high-column-density regions of five CHVCs. 

\item An old population cannot exceed several percent of the stars detected
(1 to 5\%) in the same fields. 

\item It is highly unlikely that all five CHVCs contain stellar populations which all went
undetected by chance; but there is a fifty-fifty chance for obtaining our result
if the fraction of CHVCs with stars is less than 13\%.

\end{enumerate}

\section{2MASS Data}

Having ruled out a faint population in the central regions of the
clouds with the VLT data, we now consider the possibility of more extended, nearby,
and brighter populations. To this end we make use of near-infrared photometry in the public domain.

Two Micron All Sky Survey data\footnote{The 2MASS project is a collaboration between The University 
of Massachusetts and the Infrared Processing and Analysis Center (JPL/Caltech). Funding is provided 
primarily by NASA and the NSF.} were extracted via the Infrared Processing and Analysis Center (IPAC) 
WWW site. 2MASS has finished uniformly scanning the entire sky in three near-infrared bands, J, H, and K$_S$, 
using a pixel size of 2''. 2MASS point-source photometry achieves a 10-$\sigma$ (0.109~mag) detection 
at the following magnitude levels: J = 15.8, H = 15.1, and K$_S$ = 14.3 
for unconfused sources outside of the  Galactic plane.

The second incremental data release contains data for four of our CHVCs 
(there are no data available for HIPASS~J1712-64). A major advantage of using the 2MASS archive for our search 
for stellar content is that we can easily obtain data covering a field of view which is well matched to the 
angular size of a CHVC. To this end we extracted from the point source catalog all sources within a 1$^o$ radius of 
the centre of each CHVC (as well as for several companion galaxies of the Milky Way). A further advantage 
of the 2MASS data is that they allow us to investigate Galactic
foreground contamination. We thus also extracted data in comparison fields, at the same Galactic
latitude, b, but at longitudes of l+3$^o$, and l-5$^o$, for each cloud (no data are available for
the l-5$^o$ field of HVC039-37-231). The source count in each 
of these fields is around  tenthousand, but with considerable variation. The latter  
is explained in part by the differences in area coverage within each of the 2$^o$ fields 
illustrated in Figs.~6 \& 7 (in addition to true variation due to different Galactic coordinates). 

In Figure~6, we display positional plots of the sources detected by 2MASS.
Figure~7 shows results for offset fields. There are no enhancements of stellar density
visible toward the centres of the ``ON" fields which could be interpreted as dwarf galaxies in the clouds.
The ``OFF" fields exhibit a high stellar density, showing that even at Galactic
latitudes above 30$^o$, the Galactic foreground contamination is strong in the near infrared.
The ON and OFF fields are qualitatively similar. A one-dimensional KS test on the K-band
LFs agrees with this statement (although the probabilities differ largely from case to case -- possibly 
due to small real variations with position over a few degrees in the Galaxy). The distribution of sources on the ON CMDs
is concluded to be consistent with a population of Galactic foreground stars. This leads us to favor 
the hypothesis that there are no associated stellar populations detected in these CHVCs.

In Figure~8, we show the [J-K$_S$,~K$_S$] CMDs of the data. We do not show CMDs of the OFF fields
because there is no obvious difference between the CMDs of the ON and the OFF fieds. The same
is true for other filter combinations, such as the [J-H,~H], or [H-K$_S$,~K$_S$] CMDs.
We overplotted Girardi et al. (2000) isochrones onto one of the [J-K$_S$,~K$_S$] CMDs. 
At distances of around 150~kpc (as shown in Fig.~8), we can expect to detect the full range of red 
stellar populations, supergiants, AGB stars and also, stars along the RGB. Only for distances of less 
than about 300~Kpc do the data include the TRGB. For a distance
of 1~Mpc, we could at best expect to detect red supergiants in associated galaxies with these data.
This limits the tests we can perfom with the 2MASS data. On the other hand, as shown by
Fig.~8, the 2MASS data complement the optical data in ruling out the presence of a young
stellar population. They do so even in the presence of a small amount of internal dust, 
since absorption is much smaller in the near-IR than in the optical.

The 2MASS data are a powerful tool to separate a potential very young stellar population
in HVCs from Galactic foreground stars. Notice how the 10~Myr isochrone of Fig.~8 runs
through a part of the CMD where HVC039-37-231 does not exhibit any stars. The 2MASS data
are actually sensitive to 10-Myr-old MS stars for distances of up to 125~Kpc, and for 10-Myr-old BSG
stars for distances of up to 500~Kpc. The lack of stars with colours J-K$_S$$<$0.2 demonstrates that
we do not detect a young stellar component in the CHVCs out to large distances.
 
The 2MASS data are also uniquely sensitive to young and intermediate-age AGB stars for distances 
of less than about 300~Kpc. See how old isochrones in Fig.~8 stick out of the data distribution.
They terminate at the first thermal pulse and so, do not extend as far red in colour
as the full range of AGB stars seen, e.g., in the LMC and SMC (see below). 
There are no stars with K$_S$$<$14 and J-K$_S$$>$1, where we would expect to see the
AGB component of CHVCs for distances less than about 300~Kpc. The distribution of sources in this
colour and magnitude range is instead fully consistent with Galactic foreground, only.

\subsection{CHVCs versus Local Group Dwarf Galaxies}

The CMDs of the CHVCs are quite different from 2MASS-based CMDs of the centres of
the star-forming Irregular companions of the Milky Way, the LMC and the SMC. Their
CMDs, shown in Figure~9, exhibit the blue plume near J-H, J-K$_S$ of zero. They also show a well populated
RGB, which turns into a luminous, extended AGB for J-H, J-K$_S$ colours larger than one.
The CMDs of the CHVCs do not exhibit a blue plume at colours similar to those seen
in the Magellanic Clouds. They do not show AGB stars with colours similar to those
seen in the MCs. This rules out the presence
of a young stellar content in the CHVCs with properties similar to those of Irregular
galaxies of the Local Group. 

The AGB stars are a particularly
prominent feature in the CMDs of the LMC and the SMC. Following Nikolaev \& Weinberg (2000)
the tail of AGB stars extending redward of J-K$_S$$>$1.4 is identified with carbon 
rich thermally pulsing, or TP-AGB stars, having presumed ages of 1---4~Gyr. On
CMDs in 2$^o$ fields centered on the LMC and the SMC, we find that the brightest such
AGB stars occur at K$\approx$10 in the LMC, and at K$\approx$10.5 in the SMC. Adopting the LMC \& SMC
distances given in Mateo (1998), 18.5 \& 18.8 (or 49 \& 58~Kpc), respectively, the K-band absolute
magnitudes of these C-rich TP-AGB stars are about --8.5. We would have cleary seen such
an AGB feature to within about half a magnitude of the detection limit.
Since a prominent AGB is not seen to K$\approx$14, then the CHVCs are
either more distant than m-M=22.5 or d=0.3~Mpc, or, if they are within about
0.3~Mpc, they do not contain a population of C-rich TP-AGB stars similar to that
seen in the LMC and the SMC.

Another Milky Way companion with a well-known population of bright, and intermediate-age
AGB stars is the dSph Fornax. Saviane, Held \&
Bertelli (2000) analyzed it most recently, and emphasised the presence of bright
AGB stars, which they attribute to an intermediate-age (2--10~Gyr) stellar population.
Fornax is at quite a high Galactic latitude, of almost -66$^o$, and suffers less
from Galactic foreground contamination than do our CHVCs, the LMC and the SMC.
Fornax has a distance modulus of 20.70, or a distance of 138~Kpc, 
a V-band central surface brightness of 23.4~magnitude~arcsec$^{-2}$,
and an exponential scale length of 10\farcm2 (Mateo 1998). 

The 2MASS data of Fornax, shown in Figure~9, exhibit a population
of AGB stars with J-K$>$1, and possibly of up to 2, near K$\approx$13.5.
The C-rich TP-AGB stars with colours of J-K$_S$$>$1.4 identified in the MCs
are very sparse in Fornax. These stars
extend to the red from a more strongly populated AGB feature seen between 1.0$<$J-K$_S$$<$1.4.
The magnitudes of these AGB stars, most probably the young AGB component of
Fornax, are near K$\approx$13.5. This implies absolute K magnitudes of $\approx$--7.2. 
Since similar AGB stars are not seen to K$\approx$14 in the CHVCs, then the CHVCs are
either more distant than m-M=21.2 or d=0.17~Mpc, or, if they are within 
0.17~Mpc, they do not contain a population of AGB stars similar to that
seen in Fornax.

We also culled data for the dSph galaxies Sextans (b$\approx$42$^o$), Sculptor (b$\approx$--83$^o$)
and Ursa Minor (b$\approx$45$^o$) from the 2MASS database. 
Data for Carina and Draco are not available in the second incremental data release. All four dSph
observed have exponential scale lengths of the order of 10~arcmin.
Fornax and Sculptor have rather high 
V-band central surface brightnesses, of 23.4 and 23.7~mag~acrsec$^{-2}$ (see
Mateo 1998), 
and they are easily seen as 
concentrations in the position plots shown in Fig.~10. The surface brightnesses Ursa Minor and
Sextans are only 25.5 and 26.2~magnitude~acrsec$^{-2}$ respectively. They are not distinguishable on
position plots against their respective Galactic foreground contaminations, 
although their H-band luminosity 
functions exhibit a feature where we would expect to find their TRGBs 
(compare Schulte-Ladbeck et al. 1999). 

\subsection{Detection Limits}

We performed KS tests using the K-band RGB LF of Sculptor (whose distance is 
about 80~Kpc, see Mateo 1998). 
We first selected stars by position to occur near the centre of the field of
view. This emphasises true member 
over Galactic foreground stars. We then constrained the colour and luminosity range for the
RGB to be 0.7$<$J-K$_S$$<$1.4, and K$_S$$>$15. The Galactic foreground of the CHVCs is
much stronger than that of Sculptor, since the former are situated at about 50$^o$ lower Galactic
latitude. We needed to add multiples of the Sculptor RGB stars to the [J-K$_S$,~K$_S$] CMDs of the
CHVCs before we could discriminate the additional RGB stars against the strong Galactic foreground of the
CHVC fields. The data start to suggest the presence of a Sculptor-like RGB when four times
the Sculptor stars are added to the data. Significant differences with probabilities of
less than 2\% for a chance event occur when about eight times the amount of Sculptor RGB 
stars is added to the data. A factor of eight in flux corresponds to about 2.3 mag. 
Therefore, in order to detect a Sculptor-like
RGB population in our CVHC fields, it would have to have a V-band surface brightness of
about 21.4~magnitude~acrsec$^{-2}$. This is much higher than exhibited by any of the dSph or dIrr/dSph
companions of the Milky Way. We note that M~31, however, has such high-surface-brightness
companions, e.g., NGC~147.

The 2MASS data allow us to set additional limits 
on the absence of a dwarf-galaxy like stellar population for four CHVCs. 

\begin{enumerate}

\item The distribution of stars on the sky in 2$^o$ fields centered on four CHVCs is similar
to the distribution of stars on the sky at the same Galactic latitude and a few degrees away
in Galactic longitude. A KS test cannot distinguish the K-band LFs of the ON from the OFF fields.

\item 2MASS data are sensitive to young MS stars out to about 125~Kpc and to BSGs out to
500~Kpc. None are seen. Any luminous young stellar population contributes less
than a few percent of the stars. 

\item 2MASS data are sensitive to AGB stars out to about 300~Kpc. None are seen. 
Any luminous AGB stellar population contributes less than a few percent of the stars. 

\item An old RGB population needs to have a high surface brightness to be
distinguishable against high Galactic foreground at low latitudes. 
Such a high-surface-brightness galaxy 
would be recognized in the optical on Palomar Observatory Sky Survey (POSS) plates
(compare Simon \& Blitz 2002).

\end{enumerate}

\section{Conclusions}

We conducted a two-pronged search for stellar content in five HVCs. Our optical
data are sensitive to the presence of RGB stars for distances of up to 2~Mpc, i.e., throughout
the entire Local Group. The near-infrared data are uniquely sensitive to luminous AGB stars 
with distances of up to 300~Kpc. 

We show that the optical data rule out the presence of young, blue stars in the small,
6\farcm8 x 6\farcm8, fields observed and centered on the highest column density
regions of five HVCs.  Our strong quantitative limits on the absence of RGB stars, the type of stellar
population seen in all Local Group dwarf galaxies, in the HVCs leads us
to conclude that it is highly likely that no star formation has occurred in them over the Hubble time.

We use 2MASS data (readily available for four of the five HVCs from the 2MASS archive) to investigate
the stellar content in 2$^o$ fields centered on the CHVCs. The near-IR CMDs are sensitive to 
reveal the presence of nearby young main-sequence, and of young and intermediate-age AGB stars. 
None are seen. We also place further quantitative limits on the RGB population.
In addition, we show that the spatial distribution of stars
in the 2MASS fields is indistinguishable from that in adjacent fields. In particular,
the CHVCs data lack a central concentration that gives away the presence of Milky-Way dSph companions
with typical scale-lengths of order 10~arc~min (but not of the lowest-surface 
brightness ones, such as, e.g., 
Ursa Minor dwarf Spheroidal). We did not try to search very large fields for extremely diffuse populations
of AGB or RGB stars.

We detect zero associated resolved stellar content in five HVCs investigated. 
The probability that all five 
HVCs contain stars which escaped us by chance is extremely small.
Our data therefore add to the accumulating evidence that the Local Group may contain
dark matter halos in which no star formation ever took place. 

We also conclude that our result of zero detections in five trials is yet consistent
with the hypothesis that 13\% of CHVCs contain stars.
A better knowledge of the true numbers of CHVCs with and without stars will 
help us to differentiate between possible scenarios, thus
serving as a guide toward improvements to models of galaxy formation. For example,
Bullock, Kravstov \& Weinberg (2001) propose that if gas accretion and star formation 
are suppressed in low-mass dark matter halos after intergalactic gas is reheated during 
reionization, then the observed dwarf satellites around our Galaxy are only those that 
assembled a large fraction of their mass before reionization and survived until today.
As a consequence, only 1\% of satellite halos would contain stars. Given the statistics 
discussed above, it is clear that our observations cannot as yet place
any constraints on a scenario in which a diminishing number of CHVCs contains stars.
This can only be accomplished by observing a larger sample.
We will present results of a wider search for resolved stars in CHVCs in paper~II. 

In light of the new results presented by Simon \& Blitz (2002), Lewis et al. (2002), and
by us, HVCs are an unlikely repository to increase significantly the dwarf-galaxy population of the Local Group 
of galaxies. Recent searches for additional Local Group dwarf
galaxies not targeted at HVCs (e.g. Whiting, Hau \& Irwin 2003) have similarly revealed that there is no large
population of Local Group dwarf galaxies of familiar type awaiting discovery. Therefore,
there appears to be a true discrepancy between the number of
dark-matter subhalos predicted in CDM models, and
the observed number of satellite galaxies.

\section*{Acknowledgments}

We thank Drs.~Appenzeller and Stahl for obtaining the Phoenix data during
a FORS2 comissioning run. The observations of the CHVCs were obtained with the 
FORS1 instrument for the observing proposal 67.B-0060(A).
This publication makes use of data products from the Two Micron All Sky Survey, 
which is a joint project of the University of Massachusetts and the Infrared 
Processing and Analysis Center, funded by the
National Aeronautics and Space Administration and the National Science Foundation.
We made extensive use of the NED database. 
RS-L thanks the Universities of Bochum and Bonn for hosting a visit sponsored
by the DFG Graduiertenkolleg 118 ``The Magellanic System, Galaxy Interaction, and the
Evolution of Dwarf Galaxies". She also acknowledges support from the
Max-Planck Institut f\"ur Extraterrestrische Physik Garching.

\bsp

\label{lastpage}

\end{document}